\documentclass[aps,prl,longbibliography,superscriptaddress,reprint]{revtex4-2}

\usepackage{graphicx}
\usepackage{amsmath}
\usepackage{multirow} 
\usepackage{physics} 
\usepackage{braket}
\usepackage{times}
\usepackage{gensymb}
\usepackage{hyperref}
\hypersetup{colorlinks=true, citecolor=blue, urlcolor=blue, linkcolor=blue}

\begin{document}


\title{Preserving multi-level quantum coherence by dynamical decoupling}

\author{Xinxing Yuan} \email{Equal contribution.}
\affiliation{CAS Key Laboratory of Microscale Magnetic Resonance and School of Physical Sciences, University of Science and Technology of China, Hefei 230026, China}
\affiliation{CAS Center for Excellence in Quantum Information and Quantum Physics, University of Science and Technology of China, Hefei 230026, China}

\author{Yue Li}\email{Equal contribution.}
\affiliation{CAS Key Laboratory of Microscale Magnetic Resonance and School of Physical Sciences, University of Science and Technology of China, Hefei 230026, China}
\affiliation{CAS Center for Excellence in Quantum Information and Quantum Physics, University of Science and Technology of China, Hefei 230026, China}

\author{Mengxiang Zhang}
\affiliation{CAS Key Laboratory of Microscale Magnetic Resonance and School of Physical Sciences, University of Science and Technology of China, Hefei 230026, China}
\affiliation{CAS Center for Excellence in Quantum Information and Quantum Physics, University of Science and Technology of China, Hefei 230026, China}

\author{Chang Liu}
\affiliation{CAS Key Laboratory of Microscale Magnetic Resonance and School of Physical Sciences, University of Science and Technology of China, Hefei 230026, China}
\affiliation{CAS Center for Excellence in Quantum Information and Quantum Physics, University of Science and Technology of China, Hefei 230026, China}

\author{Mingdong Zhu}
\affiliation{CAS Key Laboratory of Microscale Magnetic Resonance and School of Physical Sciences, University of Science and Technology of China, Hefei 230026, China}
\affiliation{CAS Center for Excellence in Quantum Information and Quantum Physics, University of Science and Technology of China, Hefei 230026, China}

\author{Xi Qin}
\affiliation{CAS Key Laboratory of Microscale Magnetic Resonance and School of Physical Sciences, University of Science and Technology of China, Hefei 230026, China}
\affiliation{CAS Center for Excellence in Quantum Information and Quantum Physics, University of Science and Technology of China, Hefei 230026, China}

\author{Nikolay V. Vitanov}
\affiliation{Department of Physics, St Kliment Ohridski University of Sofia, 5 James Bourchier Blvd, 1164 Sofia, Bulgaria}

\author{Yiheng Lin} \email{yiheng@ustc.edu.cn}
\affiliation{CAS Key Laboratory of Microscale Magnetic Resonance and School of Physical Sciences, University of Science and Technology of China, Hefei 230026, China}
\affiliation{CAS Center for Excellence in Quantum Information and Quantum Physics, University of Science and Technology of China, Hefei 230026, China}

\author{Jiangfeng Du} \email{djf@ustc.edu.cn}
\affiliation{CAS Key Laboratory of Microscale Magnetic Resonance and School of Physical Sciences, University of Science and Technology of China, Hefei 230026, China}
\affiliation{CAS Center for Excellence in Quantum Information and Quantum Physics, University of Science and Technology of China, Hefei 230026, China}

\begin{abstract}
Quantum  information processing with multi-level systems (qudits) provides additional features and applications than the two-level systems. However, qudits are more prone to dephasing and dynamical decoupling for qudits has never been experimentally demonstrated. Here, as a proof-of-principle demonstration, we experimentally apply dynamical decoupling to protect superpositions with three levels of a trapped $^9\rm{Be}^+$ ion from ambient noisy magnetic field, prolonging coherence by up to approximately an order of magnitude. Our demonstration, straightforwardly scalable to more levels, may open up a path toward long coherence quantum memory, metrology and information processing with qudits.

\end{abstract}

\date{\today}

\maketitle  

In recent years, there has been a growing interest in qutrits --- three-state quantum systems --- and generally in qudits, quantum systems with $d$ states.
Indeed, we are witnessing the early signs of breaking the binary paradigm of quantum computing by exploiting the advantages of qudits over qubits.
Qudits may indeed enable quantum processors to surpass the capabilities of classical computers even for moderately large numbers of quantum particles.
Qudits allow us to exploit the hitherto unused potential of the natural multi-level structure of the carriers of quantum information in some physical platforms, e.g. trapped ions and atoms, to design native qudit quantum computing and quantum simulation.
Moreover, some of the most spectacular applications of quantum technologies, such as simulations of lattice gauge theories \cite{Banuls2020}, quantum chemistry \cite{MacDonell2021}, and interacting spin systems \cite{Sawaya2020} are naturally formulated in terms of multilevel quantum systems.
Treating such problems with qubit quantum processors brings additional overheads in terms of number of qubits and quantum gate operations, increasing experimental difficulties.
On the contrary, well designed qudits can perform these tasks naturally.

The most obvious advantage of qudits is the exponential increase of the computational Hilbert space compared to qubits by a factor $(\frac d2)^n$, of course, at the expense of more demanding control.
To this end, it has been shown that qutrits present the ``sweet spot'' of the optimization task of the Hilbert-space dimensionality vs. control complexity \cite{Greentree2004}.
Even more importantly, qudits allow to 
simplify quantum circuits (known as qudit catalysis) \cite{Lanyon2008},
design new types of quantum protocols \cite{Bruckner2002,Molina-Terriza2005},
more complex entangled states \cite{Vaziri2002},
   greater violations of nonlocality \cite{Kaszlikowski2000},
     Bell inequalities resistant to noise \cite{Collins2002},
      qudit cluster states \cite{Zhou2003}, etc.
Qudits also present increased robustness to noise and relaxed requirements for quantum error correction \cite{Campbell2012,Duclos-Cianci2013,Anwar2014,Watson2015},
more efficient magic state distillation \cite{Campbell2014},
and even enhanced quantum sensing \cite{Ciampini2016,Shlyakhov2018}.
Entangled photonic qudits offer enhanced security \cite{Bechmann-Pasquinucci2000,Cerf2002},
higher communication capacity and robustness compared to qubits in quantum communication and quantum networks \cite{Walborn2006,Erhard2018,Cozzolino2019}, and even some fundamental advantages in multi-party communication \cite{Keet2010}.
Moreover, efficient recipes have been proposed for the most general unitary transformations of qutrits \cite{Klimov2003,Ivanov2006,Vitanov2012,Wang2020},
and concepts of universal qudit quantum processor with trapped ions have been presented very recently \cite{low_practical_2020, Rindbauer2021}. Experimental techniques are also being developed rapidly, with a scalable proposal \cite{low_practical_2020}, certification of multilevel cohererence \cite{Martin_Certification_2018}, and demonstrations of multi-qudit gates \cite{Rindbauer2021}.

The advantages of the qudits are heavily dependent on maintaining their coherence for sufficiently long time.
However, the coherence time of a qudit is determined by its transition with the shortest coherence time among all $d(d-1)/2$ transitions between its $d$ states.
Hence decoherence can be a more severe challenge for qudits compared to qubits and increasing the coherence time is even more important.

Coherence for qubits can be prolonged up to orders of magnitude to-date with dynamical decoupling, by techniques such as CPMG (Carr, Purcell, Meiboom, Gill) \cite{CPEffects, MGModified}, KDD (Knill Dynamical Decoupling) \cite{PhysRevLett.106.240501},  UDD (Uhrig Dynamical Decoupling) \cite{PhysRevLett.98.100504, PhysRevB.77.174509}, etc, as reviewed in \cite{suter_colloquium_2016} and applied in various platforms. These techniques are specifically designed for qubits, e.g. two-state atoms.
The extension of these techniques to qudits is not straightforward since a multistate system cannot be considered simply as a collection of two-state transitions, because while a given two-state transition is rephased, all other transitions dephase. Hence care must be taken to rephase all transitions simultaneously, or 
in a well controlled manner.
Recently, dynamical decoupling in general multi-level systems has been  proposed, either with pulsed  \cite{vitanov_dynamical_2015} or with continuous application \cite{napolitano_protecting_2021}. Here we adopt the scheme in \cite{vitanov_dynamical_2015} and experimentally demonstrate it in three hyperfine ground states of a trapped $^9$Be$^+$ ion. We repeatedly apply the sequence and find the coherence time of various superposed states can be prolonged up to an order of magnitude.

\begin{figure}[tb]

    \centering
    \includegraphics[width=8.6cm]{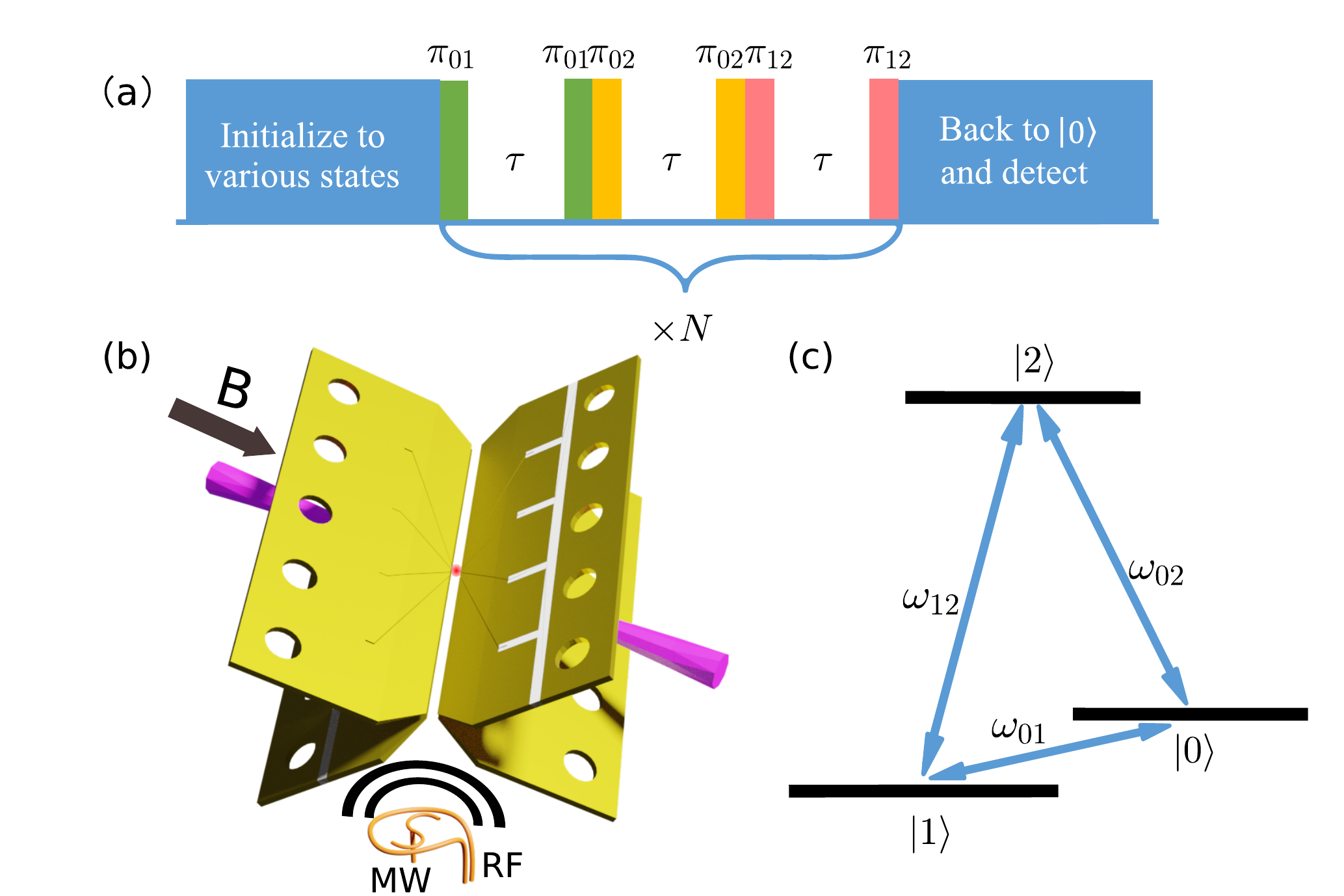}
    \caption{Scheme of the dynamical decoupling sequence and the experimental setup. (a) Scheme of dynamical decoupling sequence. The total sequence consists of $N$ repetitions of the basic unit, which includes 3 pairs of $\pi-$pulses, denoted as $\pi_{ij}$ for transition $|i\rangle\leftrightarrow|j\rangle$, separated by duration $\tau$. During the experiment, we prepare the superposition states from $|0\rangle$ and apply the dynamical decoupling sequence with various $\tau$. After the sequence, we measure the fidelity between the final state and the prepared state by reversing the preparation sequence and detect the fidelity of $|0\rangle$. (b) The ion trap system with segmented blades. The Radio Frequency (RF) blades are made of a whole blades, the DC blades have 5 segments to fine tune the potential and ion position. The RF and Microwave (MW) antenna drive the transition around 100 MHz and 1 GHz respectively. (c) The energy levels and transitions used for the demonstration of the three level dynamical decoupling sequence.}
    \label{fig:exp_setup}
    
\end{figure}

In this work, we consider a qutrit system that consists of states $\ket{i}$ with $i=\{0,1,2\}$, where transitions between each pair of states can be driven individually given sufficient spectral separations. The noise is modeled with the Hamiltonian $M=\beta\sum_i\delta_i\ket{i}\bra{i}$  (hereafter we set $\hbar=1$), where  $\beta$ is the strength of magnetic field noise and $\delta_i$ denotes the sensitivity of ambient fields for each energy level. In the general case, $\delta_i$'s are not equal to each other and their accumulation over time would lead to stochastic imbalanced phases in the superpositions of $\ket{i}$, resulting in decoherence. A technique to tackle this problem is multilevel dynamical decoupling \cite{vitanov_dynamical_2015}, where the pulse sequence $\pi_{lm} - \tau - \pi_{lm}\pi_{mn} - \tau - \pi_{mn}\pi_{nl} - \tau - \pi_{nl}$, hereafter referred to as MLDD, is repeatedly applied for $N$ times. Here $\pi_{ij}$ denotes the $\pi$-pulse driving transition $\ket{i}\leftrightarrow\ket{j}$ leading to a population exchange, $\{l,m,n\}$ are not equal to each other, and a pulse delay of $\tau$ is inserted between the pulses, as shown in Fig.~\ref{fig:exp_setup}a. 
Assuming the decoherence is mainly from the trial-to-trial variation of $\beta$ while time dependence of $\beta$ is much slower than the duration of a single trial, the main idea of MLDD is that during the sequence the population is swapped among the states, evenly experiencing $\delta_i$ and thus averaging the imbalance. 
Such an effect can also be understood by considering the overall unitary operator of MLDD. With $\pi_{lm}=e^{-i\pi(\ket{l}\bra{m}+\ket{m}\bra{l})/2}$ mentioned above, and the unitary during the wait $\tau$ as $U_M=\sum_ie^{-i\delta_i \beta\tau}\ket{i}\bra{i}$ given the slow varying assumption of $\beta$, we observe for example that $\pi_{02}U_M\pi_{02}\pi_{12}U_M\pi_{12}\pi_{01}U_M\pi_{01}=e^{-i(\delta_0+\delta_1+\delta_2)\beta\tau}I$, with $I$ an identity operator. Since the global phase is factored out from the superposition and not detectable, the application of MLDD sequence removes the differential sensitivities between the states. Deviation of the assumption of slow $\beta$ causes imperfect rephasing. Nevertheless, we analyze such effect below and show that our demonstration following \cite{vitanov_dynamical_2015} is sufficient to bring in significant improvement in coherence.

We demonstrate the dynamical decoupling sequence to protect the coherence of qutrit states in a $^9\rm{Be}^+$ ion. As shown in Fig.~ \ref{fig:exp_setup}b, the ion is placed in a Paul trap, with 313 nm lasers for cooling, state preparation and detection, and an ambient magnetic field at 13.23 mT from a pair of $\rm{Sm}_2\rm{Co}_{17}$ permanent magnets. The ground states of $^9\rm{Be}^+$ are from the coupling of orbit spin $L=0$, electronic spin $S=1/2$ and nuclear spin $I = 3/2$, leading to 8 different levels labeled by $|F,m_F\rangle$, where $F=\{1, 2\}$ is the total spin angular momentum, $m_F = \{-F,-F+1,...,F\}$ is the projection of total spin angular momentum along the quantization axis aligned with the magnetic field, see supplemental material \cite{sm}.  We label $\{\ket{2,2},\ket{2,1},\ket{1,1}\}\equiv\{\ket{0},\ket{1},\ket{2}\}$, and the transition frequencies $\omega_{ij}$ for $\ket{i}\leftrightarrow\ket{j}$ are $\{\omega_{01},\omega_{02},\omega_{12}\}\equiv2\pi\times\{116.293,995.804,1112.097\}$ MHz, respectively. The transitions are driven by applying radio-frequency (RF) and microwave fields via two separate impedance matching antennas, and resonant $\pi-$pulses are obtained with durations of approximately 25.11 $\rm {\mu s}$, 16.32 $\rm {\mu s}$ and 21.95 $\rm {\mu s}$, respectively. 

\begin{figure}[tb]
    \centering
    \includegraphics[width=8.6cm]{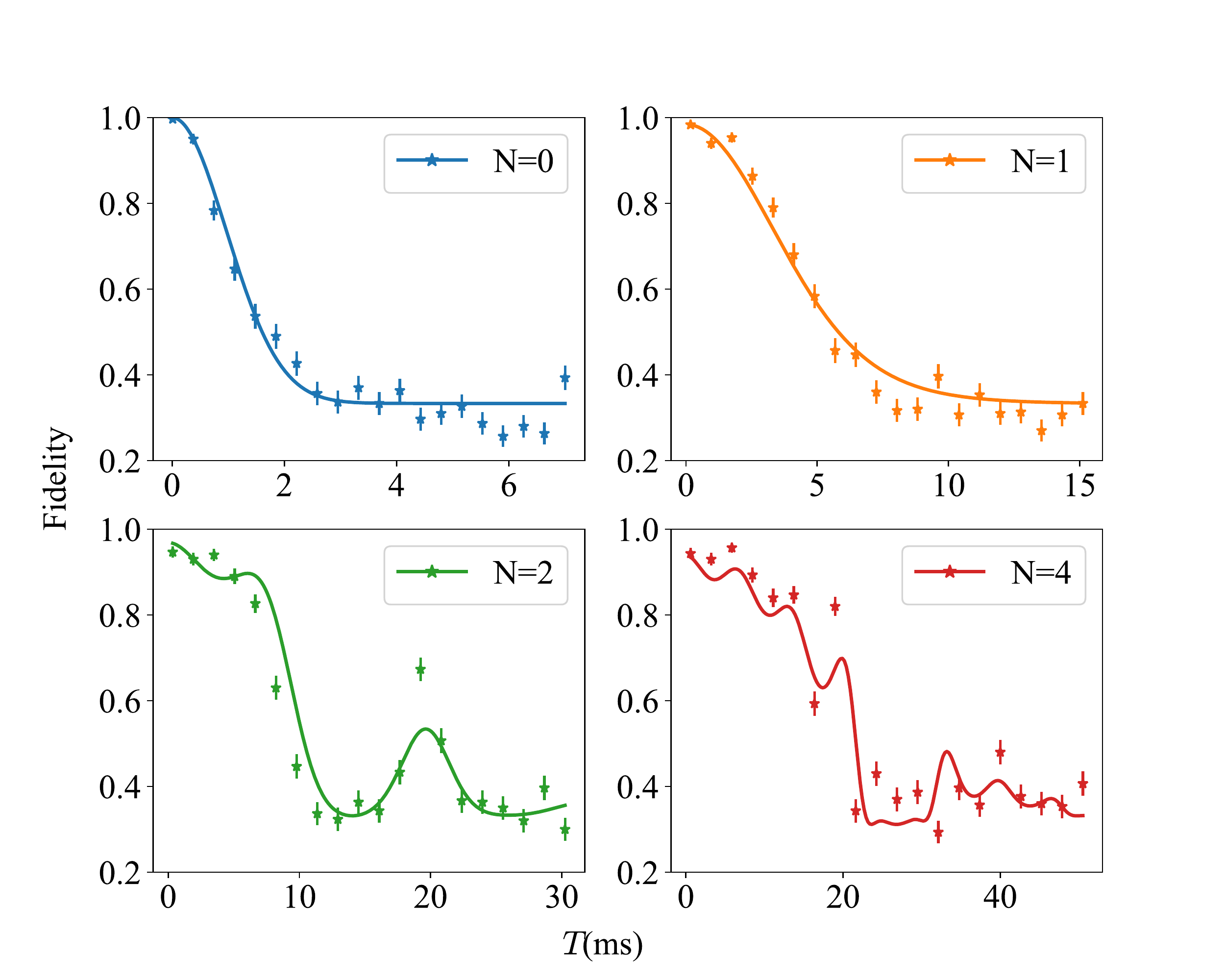}
    \caption{ The fidelity (population) of the state after the MLDD sequence. $N=0$ is the data without the protection of MLDD sequence. $N=1,2,4$ are the results of the sequences by repeating MLDD $N$ times. Each data point is obtained from 300 trials, and the error bar is estimated by binomial distribution of the raw data. The solid lines are the fit result in the framework of multilevel dynamical decoupling explained in the main text.}
    \label{fig:result1_subfig}
    
\end{figure}

The experiment begins with Doppler cooling and pumping the ion to $\ket{0}$ with a series of laser pulses. The initial superposition state $\ket{\psi_0}$ is prepared by a combination of RF and microwave pulses from $\ket{0}$, for example $\ket{\psi_0}= (\ket{0}+\ket{1}+\ket{2})/\sqrt{3}$ is prepared via an intermediate state $(\sqrt{2}\ket{0}+\ket{1})/\sqrt{3}$. The goal is to retain the phase of such superposition in $\ket{\psi_0}$  despite fluctuation of ambient magnetic field environment. After the application of MLDD for $N$ repetitions with overall duration $T$, we reverse the process of initial state preparation and map the state back to $\ket{0}$ for fluorescent detection, while the other states are mapped to states $\ket{1,-1}$ and $\ket{2,-1}$ and appear dark. 
For 400 $\rm {\mu s}$ detection with a photomultiplier tube (PMT), we typically collect 33 counts for the bright state and less than 3 counts for the dark states. Thus a threshold of 8 counts distinguishes the states. We repeat 300 trials for a typical data point and obtain the fidelity retrieving $\ket{0}$, labeled as $F(T)$. To alleviate the reduction of detection counts caused by heating during the sequence with longer wait time, we apply an off-resonant pulse to recool the ion while preserving the population before the detection, as detailed in \cite{sm}. We apply various repetitions of MLDD sequence and observe the decay of population at the detection over $T$, as shown in Fig.~ \ref{fig:result1_subfig}. 
We observe a general trend that as $N$ increases, the decay becomes slower.

To quantitatively characterize the coherence time, we parametrize possible contributions for the decay curve and try finding the best fit. Firstly, we model the reduced contrast as $g_N$, due to imperfections of the 6 applied $\pi-$pulses at each of the $N$ repetitions. Secondly, remaining stochastic noise causing an exponential decay is modeled as $c(T, T_2)=e^{-(\frac{T}{T_2})^2}$ with $T_2$ the coherence time of interest. Finally, remaining noise with known discrete spectrum, particularly multiples of $50$ Hz from alternating-current (AC) power line \cite{Shlomi_nonlinear_2013, wang_single-qubit_2017}, is modeled as $d(t,\beta)$. Combining these terms, we introduce a fit function for $N$ repetitions of the MLDD 
\begin{equation}
    F(T) = c(T, T_2)g_N \left[ d(T) - \frac{1}{3} \right] + \frac{1}{3}.
\label{eq:1}
\end{equation}
In particular, for the discrete noise, we need to have a full model to quantify the effect on the outcome. We assume the final state before detection maintains an equal superposition with an accumulated phase $\phi_i$ for each state, and is represented by $\ket{\psi_f}= 1/\sqrt{3}(\sum_ie^{i\phi_i}\ket{i})$. The subsequently detected fidelity of state $\ket{0}$ gives
\begin{align}
\begin{split}
    d(T)=\frac{1}{3}&+\frac{2}{9}\{\langle \cos[\phi_0(T)-\phi_1(T)]\rangle\\
   &+\langle \cos[\phi_1(T)-\phi_2(T)]\rangle\\
   &+\langle \cos[\phi_0(T)-\phi_2(T)]\rangle\}.
\label{eq:2}
\end{split}
\end{align}
Without restricting to the assumption that $\beta$ is a constant for each trial but rather time-dependent, denoted as $\beta(t)$,  we have
\begin{equation}
\begin{split}
    \phi_i(T)&= \sum_{k=0}^{N-1} [\delta_i\int_{\frac{k}{N}T}^{(\frac{k}{N}+\frac{1}{3N})T}\beta(t)dt \\
    &+\delta_{mod(i+1,3)}\int_{(\frac{k}{N}+\frac{1}{3N})T}^{(\frac{k}{N}+\frac{2}{3N})T}\beta(t)dt \\
    &+\delta_{mod(i+2,3)}\int_{(\frac{k}{N}+\frac{2}{3N})T}^{(\frac{k}{N}+\frac{3}{3N})T}\beta(t)dt],
\end{split}
\label{eq:3}
\end{equation}
where we assume the $\pi-$pulses have negligible duration and they are applied precisely at $(\frac{j}{3N})T$ (for $j=0,1,...,N$). 
From Breit-Rabi formula \cite{Breit_rabi} we obtain $\{\delta_0,\delta_1,\delta_2\} = 2\pi\times\{14.01,3.21,-3.20\} ~\rm{MHz/mT}$ at the value of applied magnetic field. We model $\beta(t) = \sum_j \beta_j \cos(\omega_j t+\alpha_j)$ with  $\omega_j = 2\pi \times 50\times j$ Hz, where $j=1,2,3,...$ and $\alpha_j$ is randomly averaged in each repetition of the experiment, since the experiment is not synchronized with the 50 Hz AC power line. A fit of the data is shown in Fig.~\ref{fig:result1_subfig}, and we find for $N=\{0,1,2,4\}$, the coherence time $T_2$ is fitted to be $\{1.36(4), 6.1(2), 18.4(7), 28.0(6)\}$ ms, observing improvement of coherence time up to approximately an order of magnitude. From the fits we also find $g_N=0.976(1)^N$, and a single component of the discrete noise of 150 Hz with strength of $10.0(2)~\rm{nT}$ for all the data, see supplemental materials for more details \cite{sm}.

We separately measure the coherence times  for transitions with frequency of $\{\omega_{01}, \omega_{02}, \omega_{12}\}$, obtaining values of $\{1.57(6), 1.04(4), 3.1(2)\}$ ms, respectively. Thus, the raw coherence time without applying MLDD sequence ($N=0$) for three levels approximately agrees with the shorter end among these coherence times, as expected.
The data and the overall fit result is consistent with our expectation that this scheme is best suited to remove slow varying noise, however effects from noise with high enough frequency would cause oscillations of the fidelity at certain durations, similarly known for case of two-level CPMG. We expect further improvement of the coherence time can be obtained by applying magnetic field shielding \cite{ruster_long-lived_2016,wang_single_2021}, which also improves the basic fidelity of the $\pi-$pulses, applying composite $\pi-$pulses \cite{PhysRevLett.106.240501,levitt_composite_1986}, and developing more complicated sequences as an extension of KDD for additional robustness for the sequence. Ideally, we expect the coherence time can be further extended by increasing the repetition of the MLDD sequence. However, in our setup we observe a decrease of the overall fidelity due to limited gate fidelity. The result shows that with the repetition of 8 MLDD sequences, the coherence time can be further extended to 35(1) ms, as shown in supplemental material \cite{sm}.

\begin{figure}[tb]

    \centering
    \includegraphics[width=8.6cm]{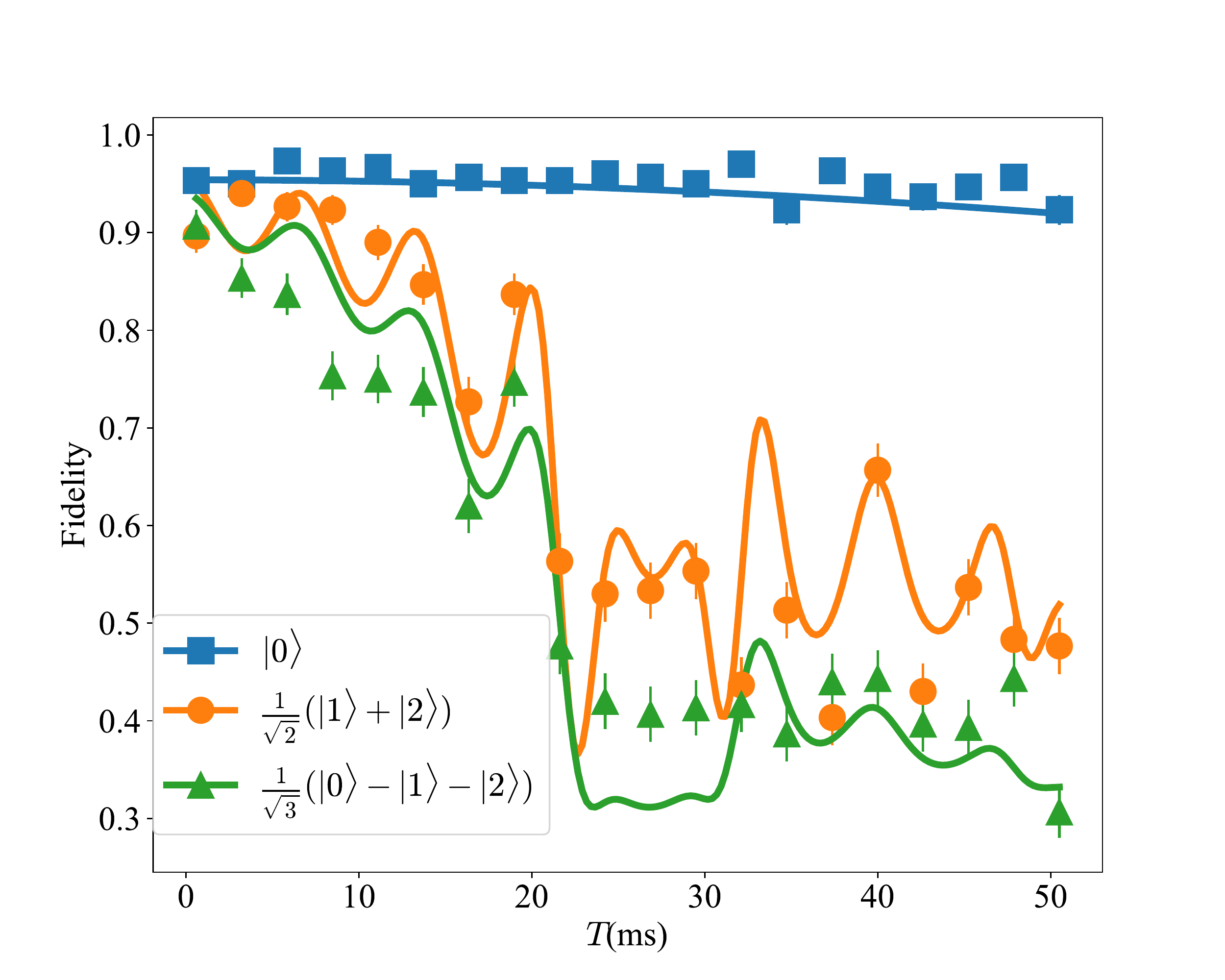}
    \caption{The fidelity of various states after the MLDD sequence with 4 repetitions, showing that such sequence can protect superpositions of two states $\frac{1}{\sqrt{2}}(|1\rangle + |2\rangle )$ and three states $\frac{1}{\sqrt{3}}(|0\rangle - |1\rangle - |2\rangle)$. MLDD results for some other superpositions are shown in \cite{sm}. The fidelity of $|0\rangle$ shows the sequence will induce about 5\% error for a single eigenstate, which remained approximately constant when varying the total duration $T$. The oscillations of the fidelity show the impact of the 150 Hz noise.}
    \label{fig:figure2}
    
\end{figure} 



To show the MLDD sequence can generally protect states in a three-level system, we prepare the initial state to various superpositions of eigenstates and measure the fidelity after 4 repetitions of the MLDD sequence, as shown in Fig.~\ref{fig:figure2}. For example, we prepare  $(|0\rangle - |1\rangle - |2\rangle)/\sqrt{3}$ and obtain coherence time of 28.0(6) ms. Similar results for other superpositions are obtained between the three levels, as detailed in \cite{sm}. We also prepare two-level superposition, such as $(|1\rangle+|2\rangle)/\sqrt{2}$, and observe the final population decays to approximately $\frac{1}{2}$ for equal mixture, in contrast to the case with the superposition of three levels decaying to approximately $\frac{1}{3}$. The theoretical prediction can be derived from Eq. \eqref{eq:1} by simply replacing the $\frac{1}{3}$ with $\frac{1}{2}$ and we can fit the result and obtain coherence times of 38(2) ms, longer than that of three-state superpositions. We attribute such effect to smaller noise sensitivity for the transition $|1\rangle \leftrightarrow |2\rangle$ than other transitions. We verify the spontaneous transition between the eigenstates is negligible and the fluorescence detection is consistent for various $T$, by preparing the state in the eigenstate of the system $\ket{0}$, and observe the fidelity hardly drops along with varying the total sequence time.

In conclusion, we implement dynamical decoupling to a three-level qutrit system of a trapped $^9\rm{Be}^+$ ion and obtain enhancement of coherence time by approximately an order of magnitude. We perform analysis of the three-level coherence decay curve, and deduce the frequency and strength for discrete noise components. 
Our results may find immediate applications in preserving the system coherence time and in quantum metrology.
Our approach can be extended to other systems, such as superconducting non-linear harmonic systems, trapped neutral atoms with multiple internal levels and nitrogen-vacancy centers. 
Since the scheme is scalable to more levels \cite{vitanov_dynamical_2015}, we expect the extension of coherence time for superpositions of even more levels to be also possible.
\\
\\
\section{Supplemental material}
\subsection{State initialization and detection}
\begin{figure}[b] 
    \centering
    \includegraphics
    [width=8.6cm]{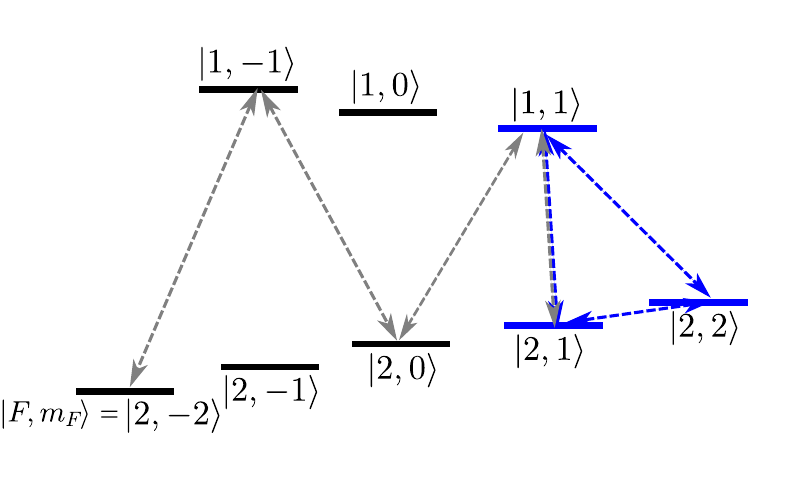}
    \caption{The energy levels of the ground state of $^9\rm{Be}^+$. The qutrit is implemented with $|2,2\rangle \equiv |0\rangle$, $|2,1\rangle \equiv |1\rangle$ and $|1,1\rangle \equiv |2\rangle$, which together with the transitions used in the sequence are colored in blue. The gray dashed transitions show the shelving path.}
    \label{fig:energy_level}
\end{figure}

\begin{figure}[t!] 
    \centering
    \includegraphics [width=8.6cm]{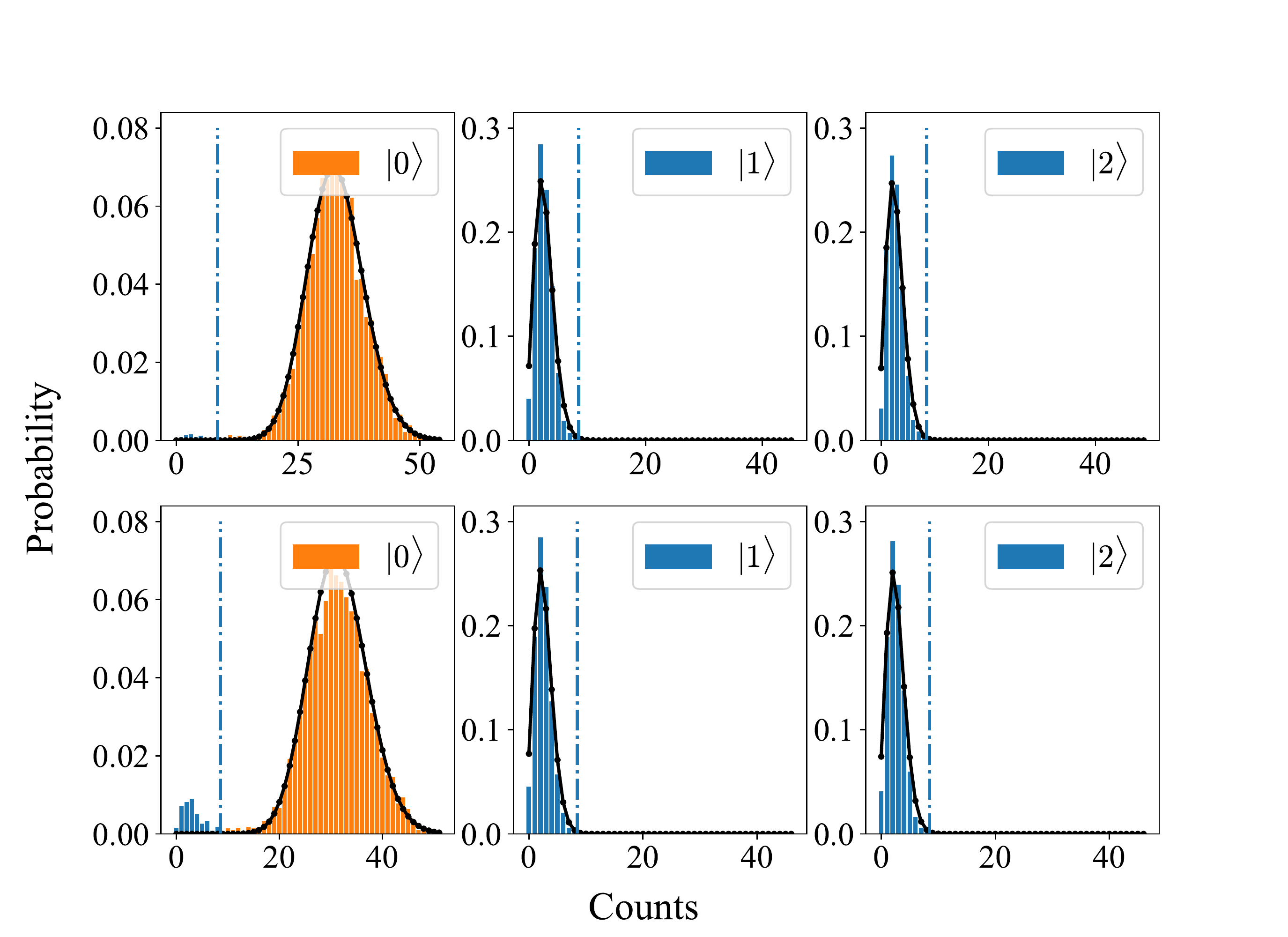}
    \caption{The detection of $|0\rangle$, $|1\rangle$ and $|2\rangle$ with pre-cooling method. The upper(lower) row shows the result of detection with 5 $\mu$s(50 ms) waiting time after the preparation with 400 $\mu$s detection time. The histogram shows the counts distribution of 5000 repetitions with counts $\leq 8$ colored in blue and counts $>8$ colored in orange. The black curves fit the counts with Poisson distribution. The detection fidelities with 5 $\mu$s (50 ms) wait time are $99.2\%$ ($96.0\%$), $99.0\%$ ($97.0\%$) and $97.4\%$ ($97.0\%$) for $|0\rangle$, $|1\rangle$ and $|2\rangle$ respectively.}
    \label{fig:detection}
\end{figure}

The ion can be initialized and Doppler cooled to $|2,2\rangle$ via a cycling transition of $|2,2\rangle \leftrightarrow |3,3\rangle$, which is also used for detection. 
As shown in Fig.~\ref{fig:energy_level}, the three blue levels and blue transitions are the ones used in the dynamical decoupling process. 
We apply optical shelving to reduce the unwanted fluorescence for the dark states to improve detection fidelity. The gray dashed transitions show the shelving path, where we transfer $|1,1\rangle$ to $|2,-2\rangle$ through $|2,0\rangle$ and $|1,-1\rangle$, and then transfer $|2,1\rangle$ to $|1,-1\rangle$ through $|1,1\rangle$ and $|2,0\rangle$.  
We find that during detection, the background heating of ion motion reduces the counts when the total sequence is longer than 40 ms, affecting the detection fidelity. 
In order to mitigate this issue, we apply Doppler laser around 10 $\rm{\mu W}$ power (the same power as  used in detection), 8 MHz red detuned from resonance, for 500 $\mu$s before detection as pre-cooling process. Our state preparation and measurement (SPAM) fidelity is more than  96.0\% with 50 ms waiting time, as shown in 
Fig.~\ref{fig:detection}. We suspect the heating is from excessive loading of ions which causes contamination to trap electrodes, since the heating effect was not so obvious in other 
 experiments in earlier stages.

\subsection{Additional coherence time measurements with dynamical decoupling sequence}
\begin{figure}[t!] 
    \centering
    \includegraphics [width=8.6cm]{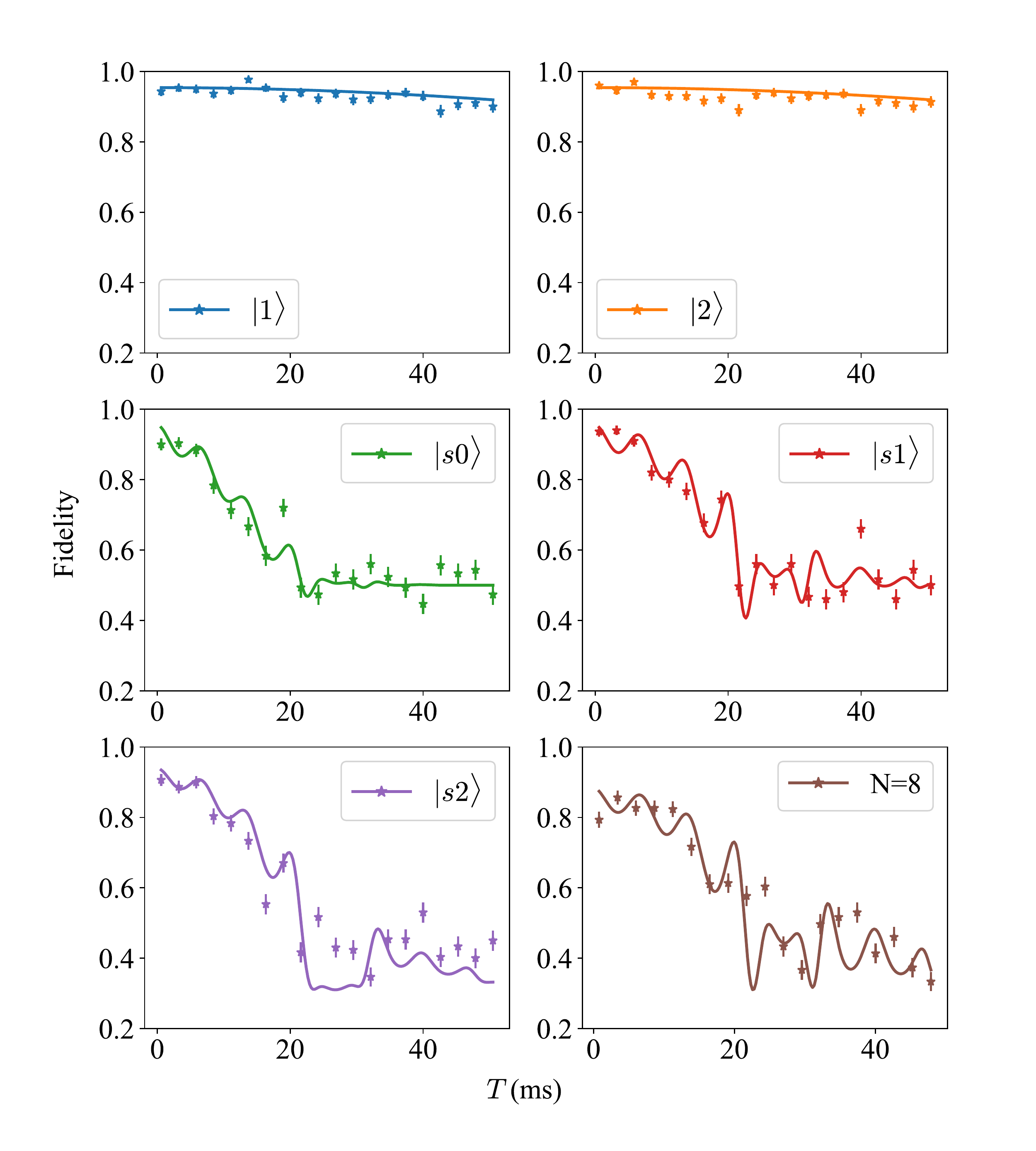}
    \caption{The effect of the MLDD sequence on various superposition states not shown in the main text. The subfigures except for the lower right one are the result of MLDD with 4 repeitions, where we obtain coherence time for $|s0\rangle \equiv \frac{1}{\sqrt{2}} (|0\rangle + |2\rangle)$ with 18(1) ms coherence time, and $|s1\rangle \equiv \frac{1}{\sqrt{2}} (|0\rangle + |1\rangle)$ with 26(2) ms, respectively. 
    The coherence time of $|s2\rangle \equiv \frac{1}{\sqrt{3}} (|0\rangle - i|1\rangle + i|2\rangle)$ are set to be the same as $\frac{1}{\sqrt{3}} (|0\rangle + |1\rangle + |2\rangle)$ in the fit. The last subfigure shows the result of $\frac{1}{\sqrt{3}}(|0\rangle +|1\rangle + |2\rangle)$ with the repetition of MLDD to be 8, corresponding to 35(1) ms coherence time.
    }
    \label{fig:others}
\end{figure}

In addition to the results in Fig.~3 in the main text, we check other superpositions between the three states of the qutrit, shown in Fig.~\ref{fig:others}.  In particular, we check the performance of MLDD with 4 repetitions, applying to states with equal populations but with other phase combinations. We fit these data together with those in the main text by using the same parameters, with fit results shown in the main text. We observe reasonable uncertainties of the fitted parameters, indicating the MLDD sequence can be applied to various phase combinations in the superposition and obtain consistent dynamical decoupling results. We also apply 8 repetitions of MLDD sequence to equal superposition state, shown in Fig.~\ref{fig:others}.
These figures together with the figures in the main text give the fit result of the strength of magnetic field 10.0(2) nT at 150 Hz, $g_N=0.976(1)^N$ and the coherence times of various states listed in the main text.

\subsection{Characterizing coherence time and operations for two-level transitions } 

\begin{figure}[h!] 
    \centering
    \includegraphics [width=8.6cm]{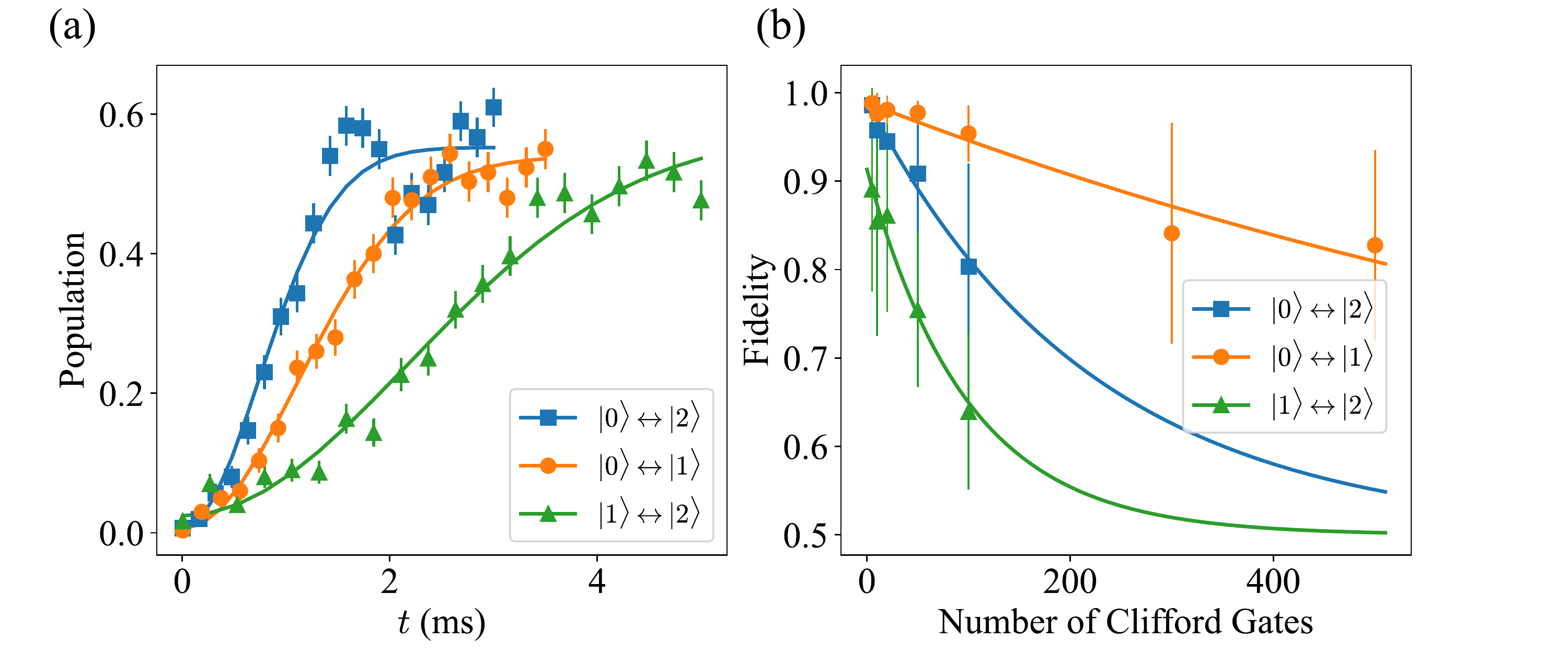}
    \caption{(a) The raw coherence time for each transition among $|0\rangle$, $|1\rangle$ and $|2\rangle$ fitted with $ae^{-(\frac{t}{T_2})^2}+b$. The fitted results give coherence times of 1.04(4) ms, 1.57(6) ms and 3.1(2) ms for $|0\rangle \leftrightarrow |2\rangle$, $|0\rangle \leftrightarrow |1\rangle$ and $|1\rangle \leftrightarrow |2\rangle$ respectively. The coherence times correspond to the respective sensitivites of the transitions. (b) The randomized benchmarking results of each transition among $|0\rangle$, $|1\rangle$ and $|2\rangle$. The average operation fidelity are 99.8(1)\%, 99.5(3)\% and 99.95(2)\% for transitions $|0\rangle \leftrightarrow |2\rangle$ , $|1\rangle \leftrightarrow |2\rangle$ and $|0\rangle \leftrightarrow |1\rangle$ respectively.}
    \label{fig:ramseyrb}
\end{figure}

We measure the coherence time of each transition of the qutrit by Ramsey experiment, where we insert a wait time $t$ between two $\pi/2-$pulses, as shown in 
Fig.~\ref{fig:ramseyrb}. 
These results agree with the sensitivities calculated with Breit-Rabi formula, and agree with the raw coherence time of qutrit without the protection of MLDD sequence.

The MLDD sequence consists of 6 $\pi-$pulses and the infidelity of the pulses will affect the fidelity of the qutrit after the MLDD sequence. We characterize the average operation fidelity of each transition by randomized benchmarking \cite{Gaebler2012,knill_randomized_2008}. The sequence is the interleave of “Pauli gate" and “Clifford gate", the form of which are $P\equiv e^{\pm i\sigma_p\pi/2}$ and $C\equiv e^{\pm i\sigma_c \pi/4}$ respectively, where $p=\{0, x, y, z\}$ and $c=\{0, x, y\}$ with $\sigma_0$ representing the identity operator. 
For a specific sequence with 2$l$+3 length labeled with $P_{l+1}C_{l}P_{l}...C_{k}P_{k}...C_1P_1C_0P_0$ where $C_k$ and $P_k$ are the $k$-th “Clifford gate" and “Pauli gate" in sequence, $P_k$ is randomly chosen from the available values, $C_k$ are chosen randomly for $k<l$, and the last $C$ gate ($C_{l}$) is chosen so that the overall result is the eigenstate of $\sigma_z$. By measuring the final state on the $\sigma_z$ basis with various sequence length, we can fit the fidelity with $\frac{1}{2}+\frac{1}{2}(1-2\epsilon_{im})(1-2\epsilon)^l$, where $\epsilon_{im}$ is the overall infidelity from initialization, the final $C$ gate and measurement; $\epsilon$ is the infidelity of “Clifford gate" $C$ together with “Pauli gate" $P$. 
For each length $l$ (number of Clifford gates), we randomly generate 30 sequences with the same length each and repeat it 1000 times in order to get reliable results. 
The results show that the average gate fidelity of the three transitions $|0\rangle \leftrightarrow |2\rangle$ , $|1\rangle \leftrightarrow |2\rangle$ and $|0\rangle \leftrightarrow |1\rangle$ are 99.8(1)\%, 99.5(3)\% and 99.95(2)\%, respectively, see Fig.~\ref{fig:ramseyrb}. 

\section{acknowledgments}
We acknowledge support from the National Natural Science Foundation of China (grant number 92165206, 11974330),  the Chinese Academy of Sciences (Grants No. XDC07000000), and the Fundamental Research Funds for the Central Universities.
NVV acknowledges support by the European Commission's Horizon-2020 Flagship on Quantum Technologies project 820314 (MicroQC).


\end{document}